\documentclass[aps,prapplied,reprint,groupedaddress]{revtex4-2}

\usepackage{graphicx}
\usepackage{subfigure}
\usepackage{float}
\usepackage{dcolumn}
\usepackage{makecell}
\usepackage{bm}
\usepackage{upgreek}
\usepackage[colorlinks=true,linkcolor=blue,citecolor=blue,urlcolor=blue]{hyperref}

\begin{document}


\title{Compact 459~nm Cs cell optical frequency standard with $2.1\times{10}^{-13}/\sqrt{\tau}$ short-term  stability}


\author{Jianxiang Miao}
\author{Tiantian Shi}
\email{tts@pku.edu.cn}
\author{Jia Zhang}
\author{Jingbiao Chen}
\affiliation{State Key Laboratory of Advanced Optical Communication, System and Network, School of Electronics, Peking University, Beijing 100871, China}

\date{\today}

\begin{abstract}
We achieve a compact optical frequency standard with an extended cavity diode laser locked to the 459~nm 6S$_{1/2}$ - 7P$_{1/2}$ transition of thermal $^{133}$Cs atoms in a $\upphi~10~\mathrm{mm}\times50~\mathrm{mm}$ glass cell, using modulation transfer spectroscopy (MTS).
The self-estimated frequency stability of this laser is $1.4\times{10}^{-14}/\sqrt{\tau}$.
With heterodyne measurement, we verify the linewidth-narrowing effect of MTS locking and measure the frequency stability of the locked laser.
The linewidth of each laser is reduced from the free-running 69.6~kHz to 10.3~kHz after MTS stabilization, by a factor of 6.75.
The Allan deviation measured via beat detection is $2.1\times{10}^{-13}/\sqrt{\tau}$ for each MTS-stabilized laser. In addition, we measure the hyperfine structure of the 7P$_{1/2}$ energy level based on the heterodyne measurements, and calculate the magnetic dipole constant \textit{A} of the Cs 7P$_{1/2}$ level to be 94.38(6)~MHz, which agrees well with previous measurements.
This compact optical frequency standard can also be used in other applications that require high-stability lasers, such as laser interferometry, laser cooling, geodesy, and so on.
\end{abstract}

\maketitle


\section{Introduction}

In the past decade, optical clock technology has seen rapid advances, with optical clocks based on lattice-trapped atoms and single-ion systems reaching uncertainty levels below $10^{-18}$~\cite{Bloom2014,Ushijima2015,Campbell2017,McGrew2018,PhysRevLett.123.033201,Norcia2019,Takamoto2020}. With such an unprecedented level of precision, optical clocks find applications in areas such as geodesy~\cite{Grotti2018,McGrew2018}, testing of fundamental physics~\cite{PhysRevLett.113.210802,PhysRevLett.113.210801,Takamoto2020} and gravitational wave detection~\cite{PhysRevD.94.124043}. However, state-of-the-art optical clocks are usually confined to laboratory environments, due to their bulk and complexity~\cite{RevModPhys.87.637}. For applications with less stringent requirements on clock performance, it is often advantageous to utilize optical clocks with thermal atomic reference, whose compactness and portability surpass their laboratory-bound counterparts. Taking advantage of their high operating frequency, compact optical clocks achieve frequency stability surpassing or comparable to microwave atomic clocks~\cite{Shang2017,Newman2019}.

Common experimental techniques for realizing a compact optical clock with thermal atoms in a cell as frequency reference include saturation absorption spectroscopy (SAS)~\cite{Cuneo1994,Genov2017,Strangfeld2021}, polarization spectroscopy (PS)~\cite{PhysRevLett.36.1170,Noh2012,Torrance2016}, two-photon spectroscopy~\cite{Touahri1997,Morzynski2013,PhysRevApplied.9.014019,Newman2021}, dual-frequency sub-Doppler spectroscopy (DFSDS)~\cite{Hafiz2016,PhysRevA.99.062508,Gusching2021} and modulation transfer spectroscopy (MTS)~\cite{Negnevitsky2013,Zi2017,Wu2018}.
These methods all utilize the interaction between counter-propagating lasers and thermal atoms to generate Doppler-free atomic spectra as the basis of laser stabilization. With SAS, PS and MTS, the counter-propagating geometry is used for velocity selection, so that only atoms with zero velocity component along the laser propagation axis are interrogated. In two-photon spectroscopy, since the Doppler shifts from the lasers in opposite directions cancel each other, atoms in all velocity classes can contribute to the signal. DFSDS uses the CPT effect in atoms with $\Uplambda$-type energy levels to generate a sub-Doppler absorption peak with narrow linewidth.
Here we list the stability levels reached by these approaches in related works, as shown in Table~\ref{context}.

Among these stabilization techniques based on thermal atoms in a cell, MTS has the advantages of insensitivity to background absorption and rejection of low-frequency noise~\cite{PhysRevLett.44.1251,Shirley1982,Ito2000}.
The 532~nm I$_2$ optical frequency standard with MTS locking has been employed as a wavelength standard due to its high stability~\cite{4126932,Schuldt2017}. Compared to iodine optical frequency standards, MTS systems based on alkali metal atoms has several advantages: alkali metal atomic cells can be made much smaller while retaining a similar level of signal-to-noise ratio as iodine frequency standards, and a wide range of diode lasers are available for interrogating different transitions of alkali metal atoms~\cite{Chang2019,Shang2020}.

\begin{table}[b]
	\caption{\label{context}%
		Stability levels reached by different types of compact optical frequency standards.
	}
	\begin{ruledtabular}
		\begin{tabular}{lcc}
			System type&Frequency stability&Reference\\
			\colrule\rule{0pt}{1.2em}%
			SAS, Rb, 780~nm&$1.7\times{10}^{-12}$ at 1~s&\cite{Strangfeld2021}\\
			PS, Rb, 780~nm&$1.5\times{10}^{-12}$ at 1~s&\cite{Torrance2016}\\
			Two-photon, Rb, 778~nm&$4\times{10}^{-13}/\sqrt{\tau}$&\cite{PhysRevApplied.9.014019}\\
			DFSDS, Cs, 895~nm&$1.1\times{10}^{-12}/\sqrt{\tau}$&\cite{Gusching2021}\\
			MTS, Cs, 852~nm&$2.6\times{10}^{-13}/\sqrt{\tau}$&\cite{Shang2020}\\
			MTS, Cs, 459~nm&$2.1\times{10}^{-13}/\sqrt{\tau}$&This work
		\end{tabular}
	\end{ruledtabular}
\end{table}

In this paper, we experimentally study a compact Cs optical frequency standard that is locked to the 459 nm 6S$_{1/2}$ - 7P$_{1/2}$ transition of $^{133}$Cs.
After optimizing the operating parameters of the optical frequency standard, we measure the frequency stability of the output laser via 
both self-estimation and 
heterodyne measurement between two identical MTS systems. 
The self-estimated Allan deviation is $1.4\times{10}^{-14}/\sqrt{\tau}$ and reaches a minimum of $4\times{10}^{-15}$ at 30~s averaging time.
By measuring the frequency spectrum of the beat signal between two MTS-stabilized lasers and the beat signal between one MTS-stabilized laser and free-running laser, we verify the linewidth-narrowing effect of MTS locking, with the linewidth of an individual laser reduced from 69.6~kHz to 10.3~kHz, by a factor of 6.75.
The Allan deviation of the beating signal is $3\times{10}^{-13}/\sqrt{\tau}$; assuming the two MTS systems contribute equally to the fluctuations of the beat signal, each MTS system has short-term frequency stability of $2.1\times{10}^{-13}/\sqrt{\tau}$.
We also measure the hyperfine level splitting of the 7P$_{1/2}$ level from the beat signal between two MTS systems locked to the Cs 6S$_{1/2}$ (\textit{F}=4) - 7P$_{1/2}$ (\textit{F'}=3) transition and 6S$_{1/2}$ (\textit{F}=4) - 7P$_{1/2}$ (\textit{F'}=3\&4) crossover transition, respectively, and calculate the magnetic dipole constant \textit{A} of the Cs 7P$_{1/2}$ energy level to be 94.38(6)~MHz, which agrees with previous works~\cite{RevModPhys.49.31,Gerheardt1972,Williams_2018,PhysRevA.100.042506}.
In addition to being used as a portable optical frequency standard, this compact, narrow-linewidth, high-stability laser can have various other applications, such as laser interferometry, laser cooling, geodesy, and so on.

\section{Experiment}

\begin{figure*}[htbp]
	\subfigure{\label{scheme-mts}
		\includegraphics[width=0.67\linewidth]{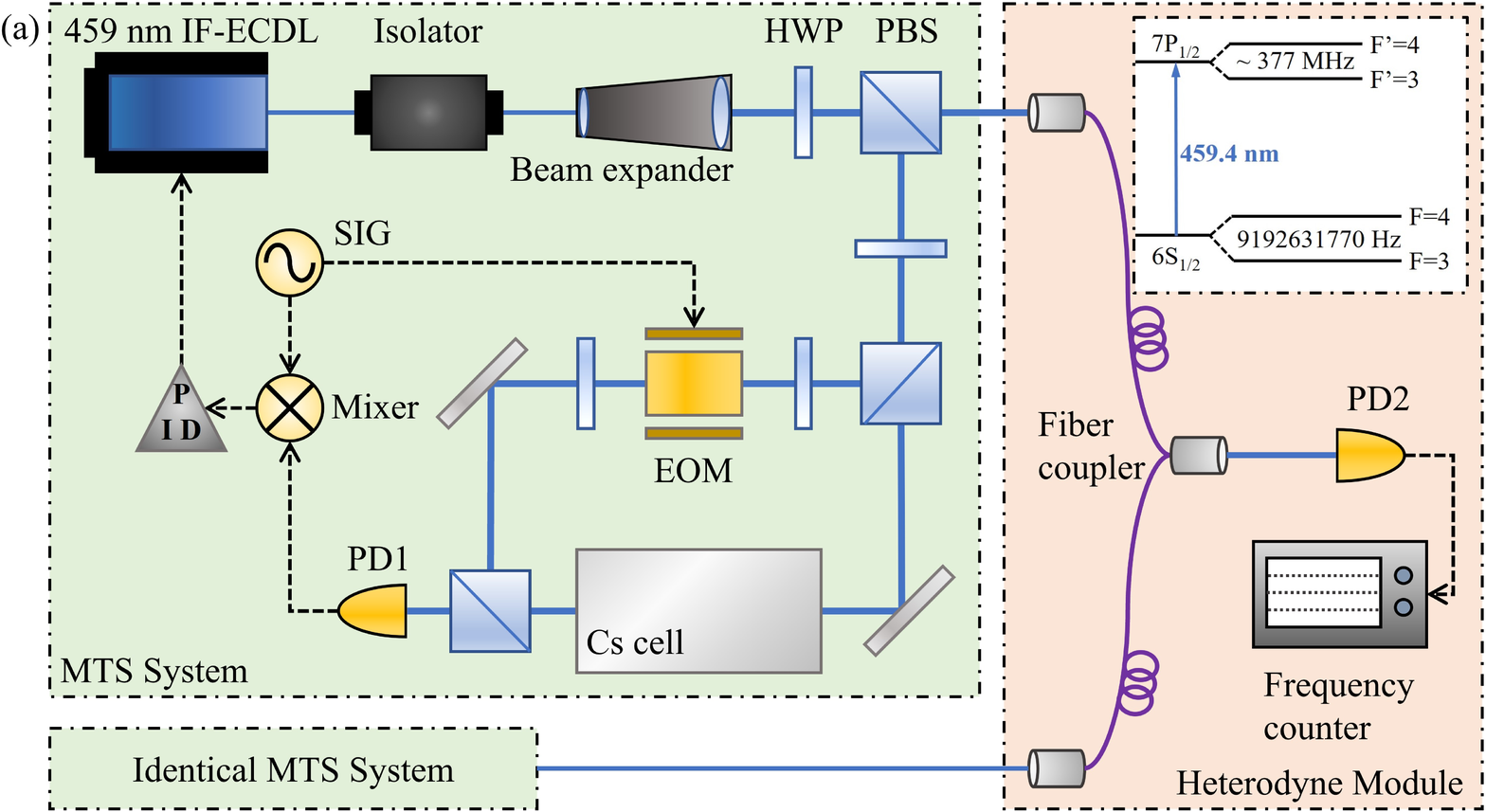}}
	\subfigure{\label{photo-mts}
		\includegraphics[width=0.3\linewidth]{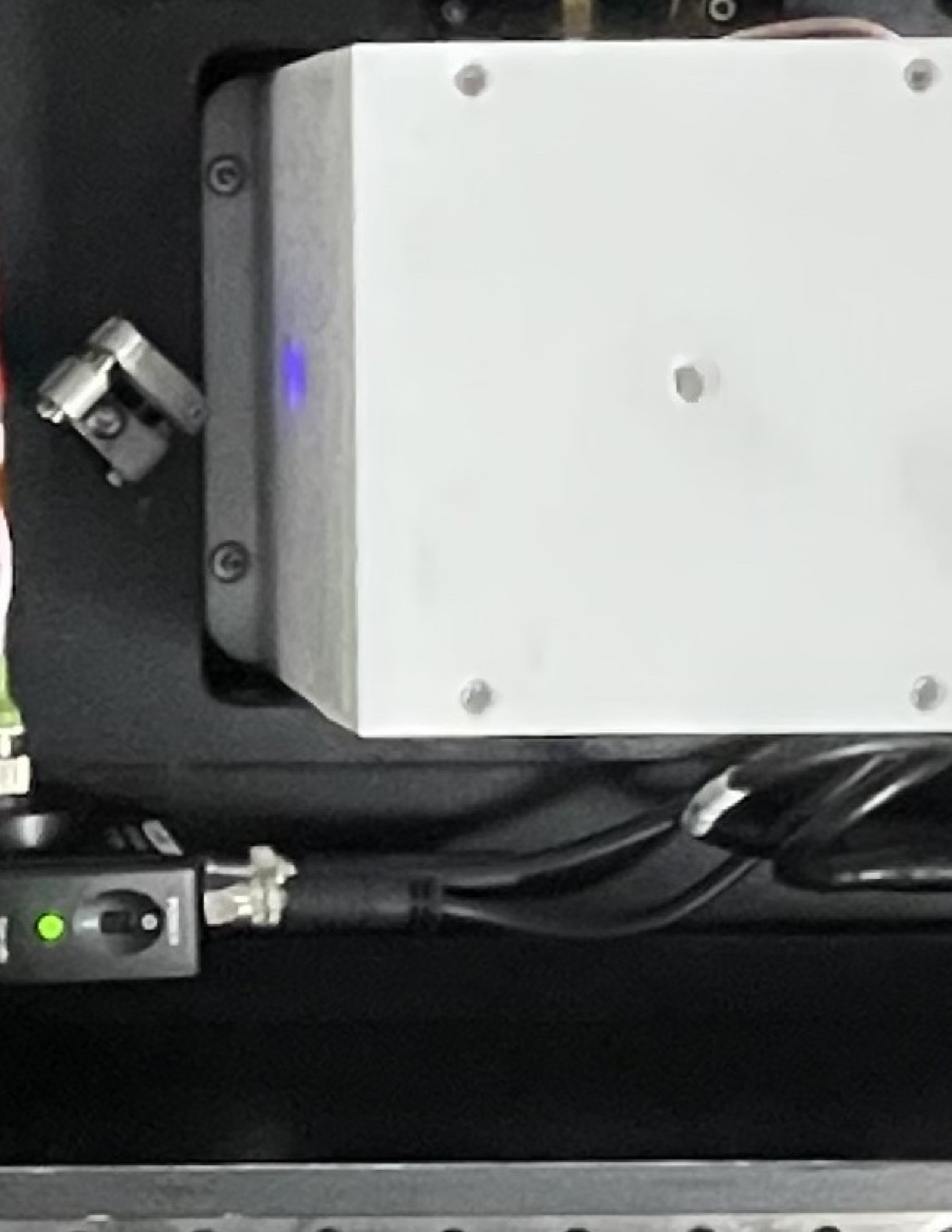}
	}
	\caption{
	(a) The experimental setup of the MTS system, with an identical MTS system set up for heterodyne measurement.
	\mbox{459~nm} IF-ECDL, \mbox{459~nm} interference filter configuration external cavity diode laser; HWP, half-wave plate; PBS, polarizing beamsplitter; PD, photodetector; SIG, signal generator; PID, proportion-integral-derivative locking system.
	Inset: hyperfine levels of the Cs 6S$_{1/2}$ and 7P$_{1/2}$ energy levels.
	(b) Photo of the experimental setup of the MTS system. All components are secured to an aluminum baseboard for better mechanical stability, and the setup is enclosed during its operation to reduce the effect of external disturbances. To the left of the photo are components irrelevant to the MTS experiment, which are therefore omitted from the photo; the MTS system can in principle be enclosed in the area prescribed by the photo, which is $44\times40~\mathrm{cm}^2$, and the box containing the experimental setup is 16~cm high. The fiber-coupled heterodyne module and the other MTS system are not shown in the photo.}
\end{figure*}

The schematic of the 459~nm laser system frequency-stabilized by MTS is shown in Fig.~\ref{scheme-mts}, and the photo of the system is shown in Fig.~\ref{photo-mts}. We use a 459~nm interference filter configuration external cavity diode laser (IF-ECDL) as local oscillator for the MTS-stabilized optical clock. The frequency of the IF-ECDL is tunable via adjusting the injection current and the voltage on a piezoelectric ceramic (PZT) controlling the cavity length. The IF-ECDL has a maximum output power up to 30~mW, and its linewidth is measured to be below 50~kHz. After passing a Faraday isolator and a beam expander, the output laser is then split in two parts after passing through a half-wave plate (HWP) and a polarizing beamsplitter (PBS), with about 2~mW of optical power for laser frequency stabilization by MTS. The light used for stabilization is further split into probe and pump beams, with optical power 0.33~mW and 1.5~mW, respectively. The probe beam passes through a $\upphi~10~\mathrm{mm}\times50~\mathrm{mm}$ cylindrical glass cell filled with $^{133}$Cs atoms, and is then detected by the photodetector (PD1, Thorlabs PDA8A2).
The cell is housed in a casing fixed to the baseboard, with several layers of Teflon for heat-insulation and two layers of cylindrical mu-metal magnetic shielding.
The pump beam is frequency modulated by an EOM, and it overlaps with the probe beam in the cesium cell at a lin$\bot$lin counter-propagating configuration. The temperature of the cesium cell is controlled by a commercial temperature controller (Thorlabs TC200), which detects the temperature of the cesium cell via a thermistor and adjusts the current sent to the heater accordingly. A HWP is placed before the EOM to align the light polarization with the main axis of the EOM crystal. The pump beam acquires sidebands from the EOM's modulation, with the $\pm1$ sidebands dominant; and after four-wave mixing processes in the cesium cell, the modulation is transferred to the probe beam. PD1 detects the beatnote between the probe beam's carrier and the two dominant sidebands. After frequency mixing the signal from PD1 with a demodulation signal with the same frequency as the modulation frequency, a dispersion-like signal is obtained at the mixer's output port. This signal is used as the error signal for the PID locking system controlling the diode laser's injection current and PZT voltage, so the laser frequency is locked to atomic transition. The MTS system occupies a space of $44\times40\times16~\mathrm{cm}^3$.

For the evaluation of the MTS-stabilized laser's frequency stability, we use the self-estimation method for a single MTS system and heterodyne measurement between two identical MTS systems. The self-estimation method evaluates the frequency stability by measuring the residual error signal after locking, and is used for individually evaluating the frequency stability of an MTS-stabilized laser. It is a commonly used method that characterizes how well the laser frequency follows the atomic transition, and reflects on the in-loop locking performance~\cite{Ito2000,Zhao_2004,Moon2004}. For heterodyne measurement, the output laser of two identical MTS systems are coupled into two input ports of a 50:50 fiber-optical coupler, and the beat signal is detected by a high-bandwidth photodetector (PD2, Hamamatsu c5658) at the output port. Two MTS systems with identical setup are locked to the 6S$_{1/2}$~({F}=4) - 7P$_{1/2}$~({F'}=3) transition and the 6S$_{1/2}$~({F}=4) - 7P$_{1/2}$~({F'}=3\&4) crossover transition separately, and the beat signal is sent into a frequency counter (Keysight 53230A), then the frequency stability can be calculated from the recorded beat frequency. We also measure the laser linewidth and the phase noise spectrum with a RF spectrum analyzer (Keysight N9000B) and a phase noise analyzer (Keysight E5052B), respectively.

\section{Results}

\subsection{Optimization of the MTS setup}

\begin{figure}[htbp]
	\subfigure{
		{\label{F=3}}
		\includegraphics[width=\linewidth]{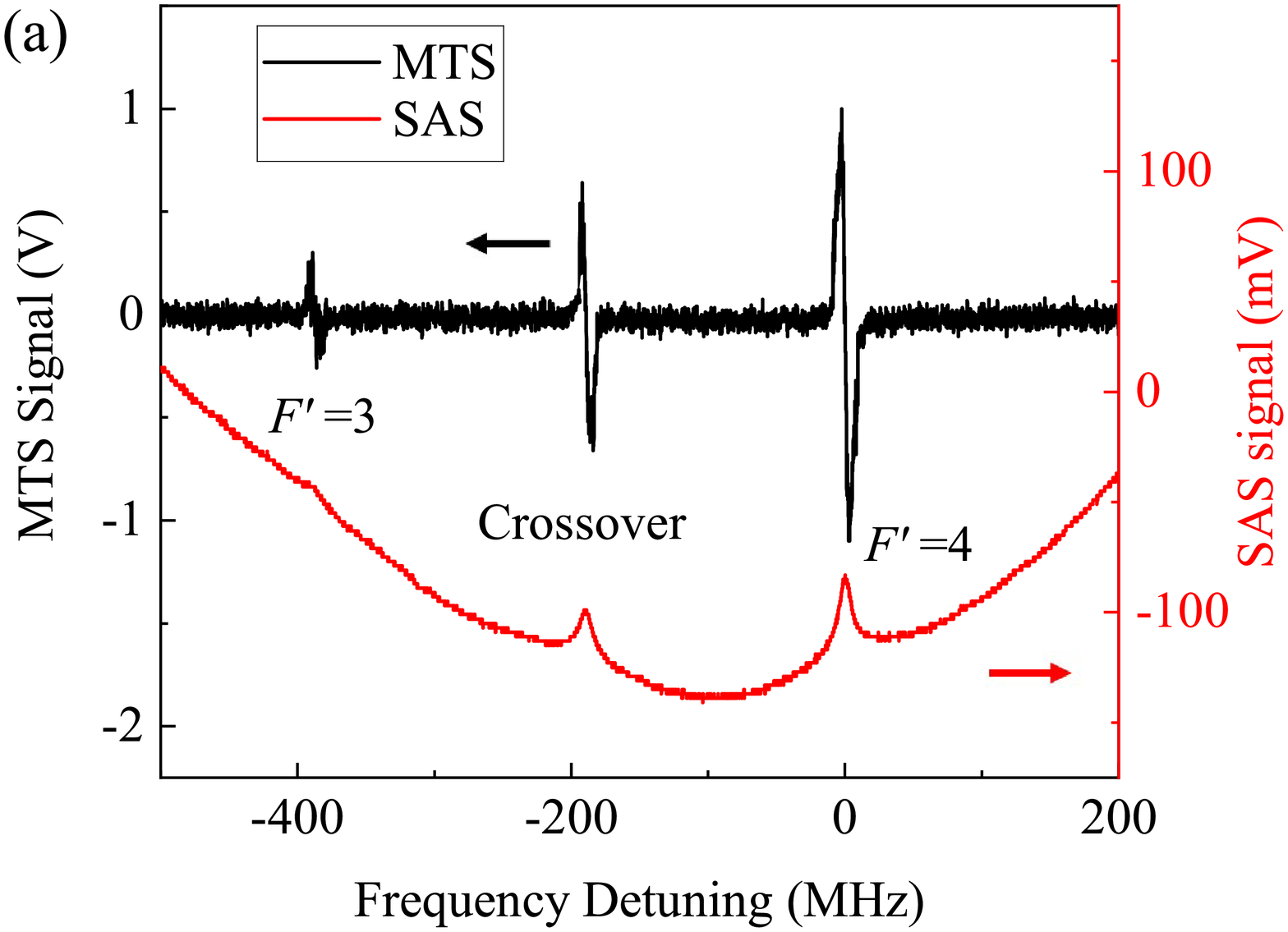}}
	\linebreak
	\subfigure{
		{\label{F=4}}
		\includegraphics[width=\linewidth]{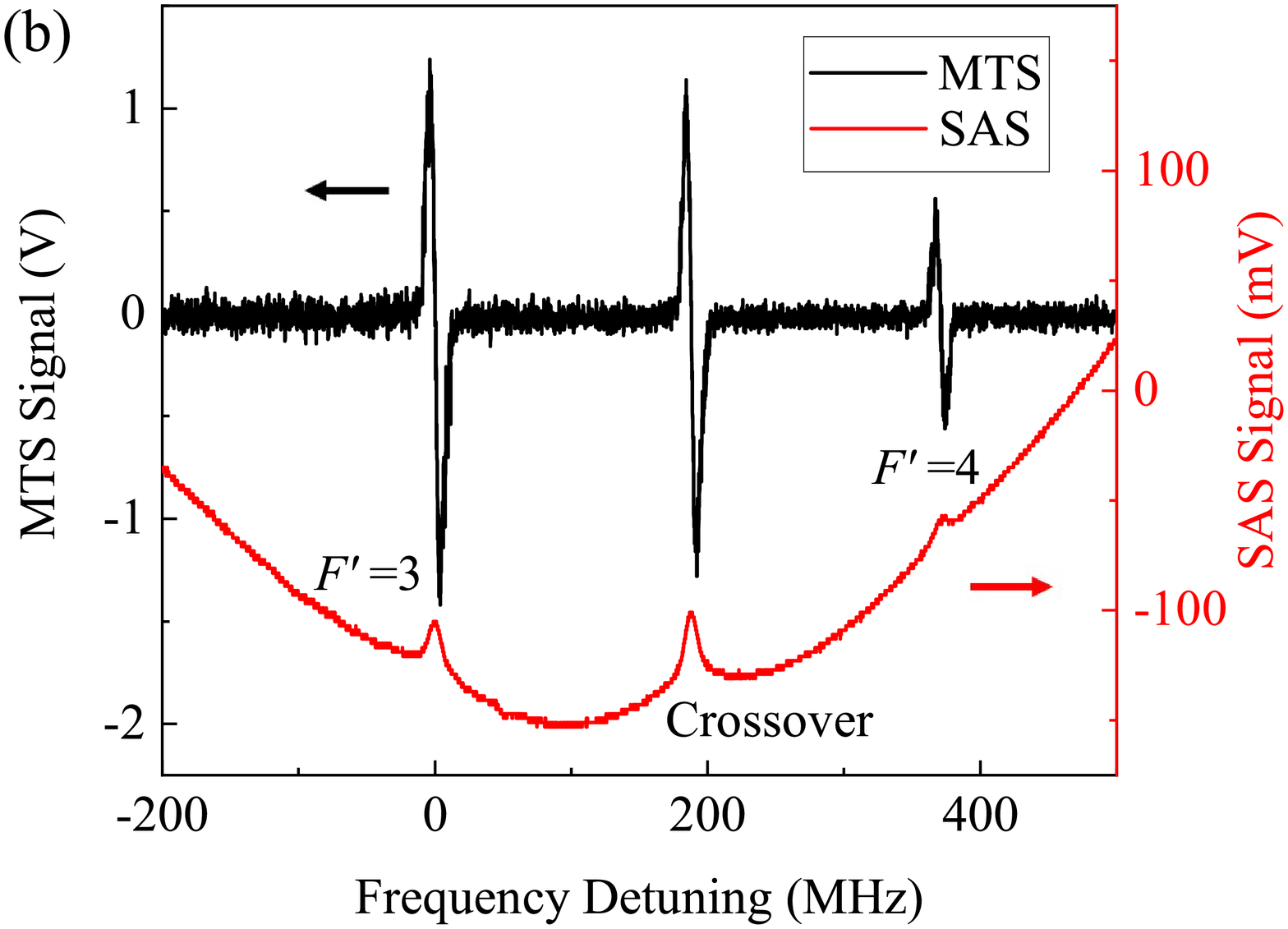}}
	\caption{SAS and MTS spectrums of the Cs 6S$_{1/2}$ - 7P$_{1/2}$ transition, starting from the (a) {F}=3 and (b) {F}=4 hyperfine levels. Comparing (a) and (b), we find the {F}=4 - {F'}=3 hyperfine transition has the largest MTS signal strength, and choose it as the reference transition. The natural linewidth of the $^{133}$Cs 6S$_{1/2}$ - 7P$_{1/2}$ transition is 6.053(7)~MHz~\cite{PhysRevA.100.052507}. The linewidth of the MTS signal corresponding to the {F}=4 - {F'}=3 hyperfine transition is measured to be 8.63~MHz from the data plot, using the hyperfine splitting (here we use the approximate value of 377.4~MHz~\cite{RevModPhys.49.31}) between F'=3 and F'=4 as a reference.}
\end{figure}

We first obtain SAS and MTS spectra of the $^{133}$Cs 6S$_{1/2}$ - 7P$_{1/2}$ transition. Figs.~\ref{F=3} and~\ref{F=4} show the MTS and SAS spectra with ground states 6S$_{1/2}$~({F}=3) and 6S$_{1/2}$~({F}=4), respectively. As can be seen from the figures, the 6S$_{1/2}$~({F}=4) - 7P$_{1/2}$~({F}=3) transition has the largest signal amplitude and slope under the same working conditions (cell temperature, EOM modulation frequency, pump and probe laser power, etc.), so we choose it as the reference transition for MTS stabilization.

\begin{figure}
	\includegraphics[width=\linewidth]{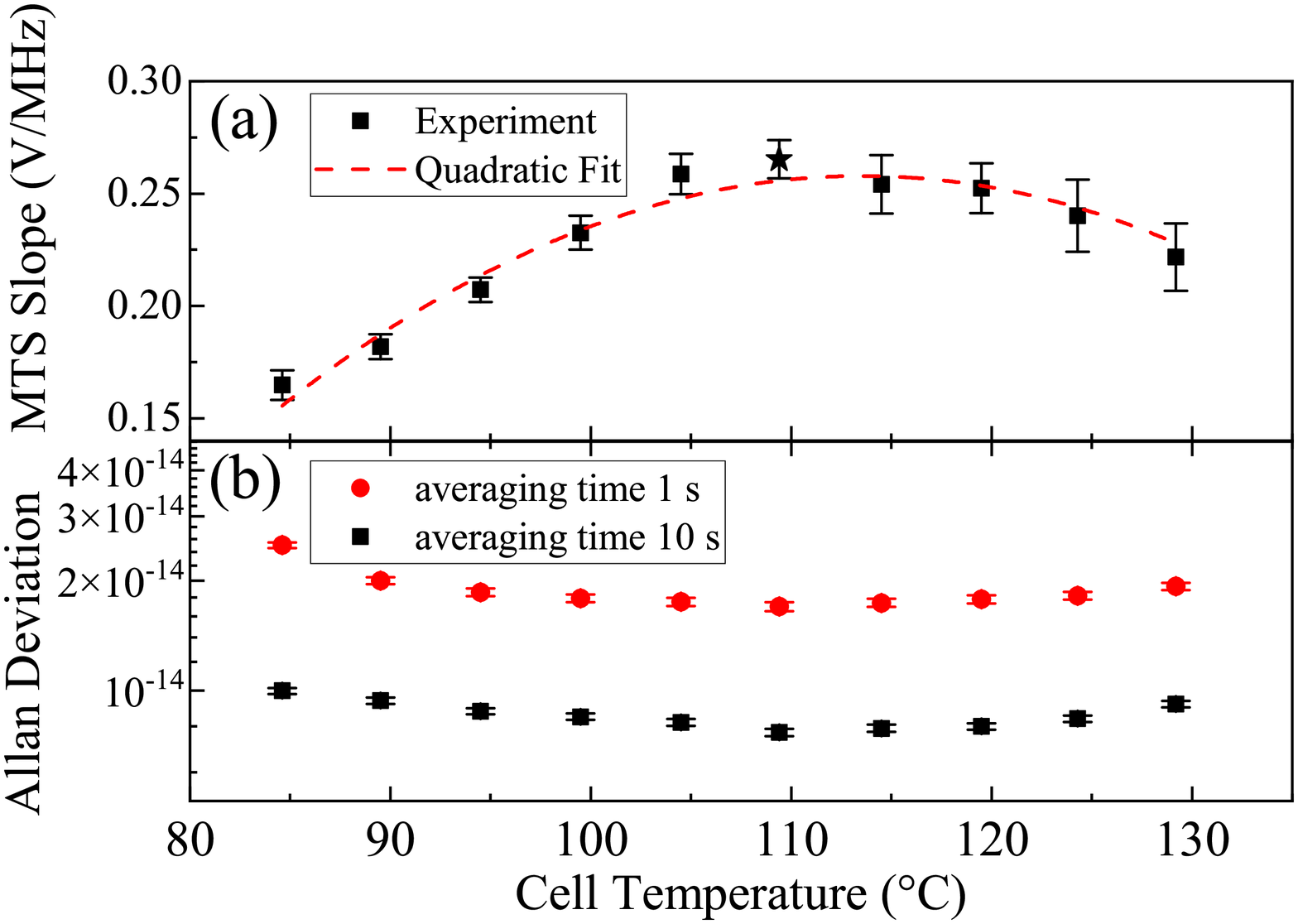}
	\caption{\label{temp}
		(a) MTS signal slope at zero crossing and (b) self-estimated short-term frequency stability, measured at different Cell temperatures, with the modulation frequency set at 1.7~MHz. The temperature we choose for the following experiments (except measurement of collision shift which changes the cell temperature) is denoted with a star symbol in (a).
		Each data point in (a) is the averaged result of 3 measurements, and the error bars represent the standard deviation.
		Each data point in (b) is obtained by monitoring the error signal for more than 100 seconds, then calculating the self-estimated Allan deviation from the recorded error signal. The error bars in (b) represent the 1-sigma confidence interval of each Allan deviation value.}
\end{figure}

\begin{figure}[htbp]
	\includegraphics[width=\linewidth]{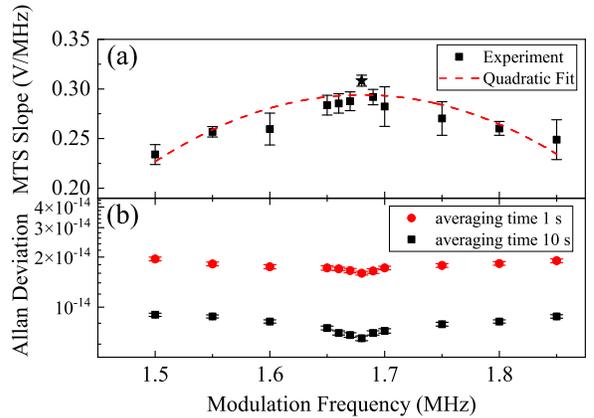}
	\caption{\label{freq}
		(a) MTS signal slope at zero crossing and (b) self-estimated short-term frequency stability, measured at different modulation frequencies, with the cell temperature controlled at 109.4$^{\circ}$C. The modulation frequency we choose for the following experiments is denoted with a star symbol in (a).
		Each data point in (a) is the averaged result of 3 measurements, and the error bars represent the standard deviation.
		Each data point in (b) is obtained by monitoring the error signal for more than 100 seconds, then calculating the self-estimated Allan deviation from the recorded error signal. The error bars in (b) represent the 1-sigma confidence interval of each Allan deviation value.}
\end{figure}

Since the laser is locked at the zero crossing of the MTS signal, its slope at the zero crossing is a good indicator of the system parameters, as a higher slope means a larger signal-to-noise ratio, as well as higher sensitivity to frequency fluctuations. We measure the MTS signal slope under various cesium cell temperatures and EOM modulation frequencies, as well as the short-term stability measured via self-estimation, in order to find the optimal working conditions. Results are shown in Figs.~\ref{temp} and~\ref{freq}. In Fig.~\ref{temp}(a), the signal slope first increases with the cell temperature due to the increase in atomic density resulting in a larger signal, then decreases as the resonance linewidth is affected by collision broadening; In Fig.~\ref{freq}(b), the signal slope increases sharply around 1.68~MHz.
The natural linewidth of the cesium 6S$_{1/2}$ - 7P$_{1/2}$ transition is $\Gamma_\mathrm{nat}=6.053(7)~\mathrm{MHz}$~\cite{PhysRevA.100.052507}, and the linewidth of the MTS signal corresponding to the 6S$_{1/2}$ ({F}=4) - 7P$_{1/2}$ ({F'}=3) hyperfine transition is measured to be $\Gamma_\mathrm{eff}=8.63~\mathrm{MHz}$ from the data plot, using the hyperfine splitting between 7P$_{1/2}$ (F'=3) and 7P$_{1/2}$ (F'=4) as a reference. This increase in the measured linewidth from the natural linewidth can mostly be attributed to saturation broadening~\cite{Ito2000}.
Here we have the modulation frequency $\omega_m=1.68~\mathrm{MHz}$ at approximately 0.2 times the measured linewidth $\Gamma_\mathrm{eff}=8.63~\mathrm{MHz}$, which can be regarded as the optimal modulation frequency for maximizing the slope of the MTS signal~\cite{Preuschoff2018}.
The short-term stability measurements in Figs.~\ref{temp}(b) and~\ref{freq}(b) also agree well with the signal slope measurements in Figs.~\ref{temp}(a) and~\ref{freq}(a). The cell temperature is set at 109.4$^{\circ}$C, and the modulation frequency is set at 1.68~MHz, both corresponding to maximal signal slope and best self-estimated short-term stability, thereby optimizing the performance of the MTS-stabilized laser.

\subsection{Characterization of MTS-stabilized laser}

\begin{figure}[htbp]
	\includegraphics[width=\linewidth]{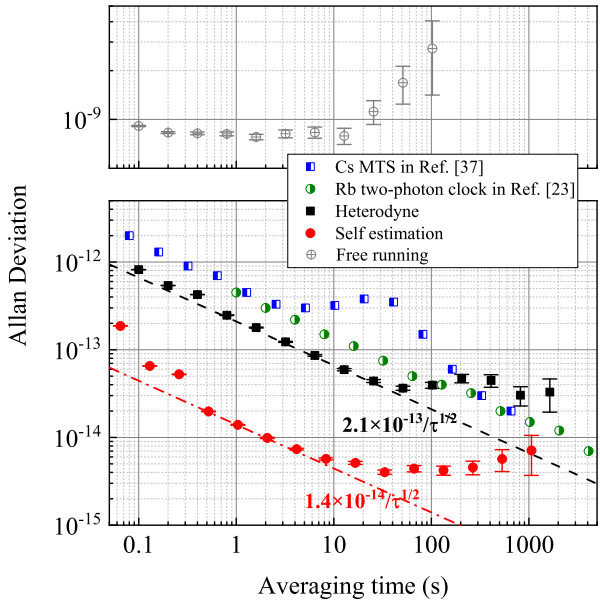}
	\caption{{\label{Allan}}Allan deviation of the 459 nm laser stabilized via MTS, measured by self-estimation (red solid circles) and heterodyne measurement (black solid squares). Compared to the frequency stability of the 852~nm laser locked to the Cs 6S$_{1/2}$ - 6P$_{3/2}$ transition in Ref.~\cite{Shang2020} (blue half-squares) and the 778~nm optical clock based on the Rb 5S$_{1/2}$ - 5D$_{5/2}$ two-photon transition in Ref.~\cite{PhysRevApplied.9.014019} (green half-circles), our MTS-stabilized laser has better short-term frequency stability at averaging time below 100 s.
	The error bars in this figure represent the 1-sigma confidence interval of each Allan deviation value.}
\end{figure}

We first use the self-estimation method to evaluate the frequency stability of the 459 nm laser after locking. The dispersion-like MTS signal is approximately linear near resonance, and the frequency shift can be derived from the residual error signal, as fluctuations of the error signal translate to frequency fluctuations. As shown in Fig.~\ref{Allan}, the short-term stability measured with self-estimation at averaging time of 1 s is $1.4\times{10}^{-14}/\sqrt{\tau}$, reaching $4\times{10}^{-15}$ at 30 s averaging time, and maintains this level of stability until after 200 s, where the Allan deviation then degrades slightly. The long-term self-estimated frequency stability is mostly limited by the residual amplitude modulation (RAM) of the EOM, and can be improved by methods such as controlling the temperature of the EOM crystal~\cite{Li2012,Zhang2014,Tai2016}. Though self-estimation is only an in-loop measurement and not truly representative of the laser frequency stability, it reflects the how well the laser frequency ``follows" the atomic transition, which sets a lower limit to the attainable frequency stability~\cite{Ito2000,Moon2004,Zhao_2004}.

\begin{figure}[htbp]
	\subfigure{
		{\label{linewidth}}
		\includegraphics[width=\linewidth]{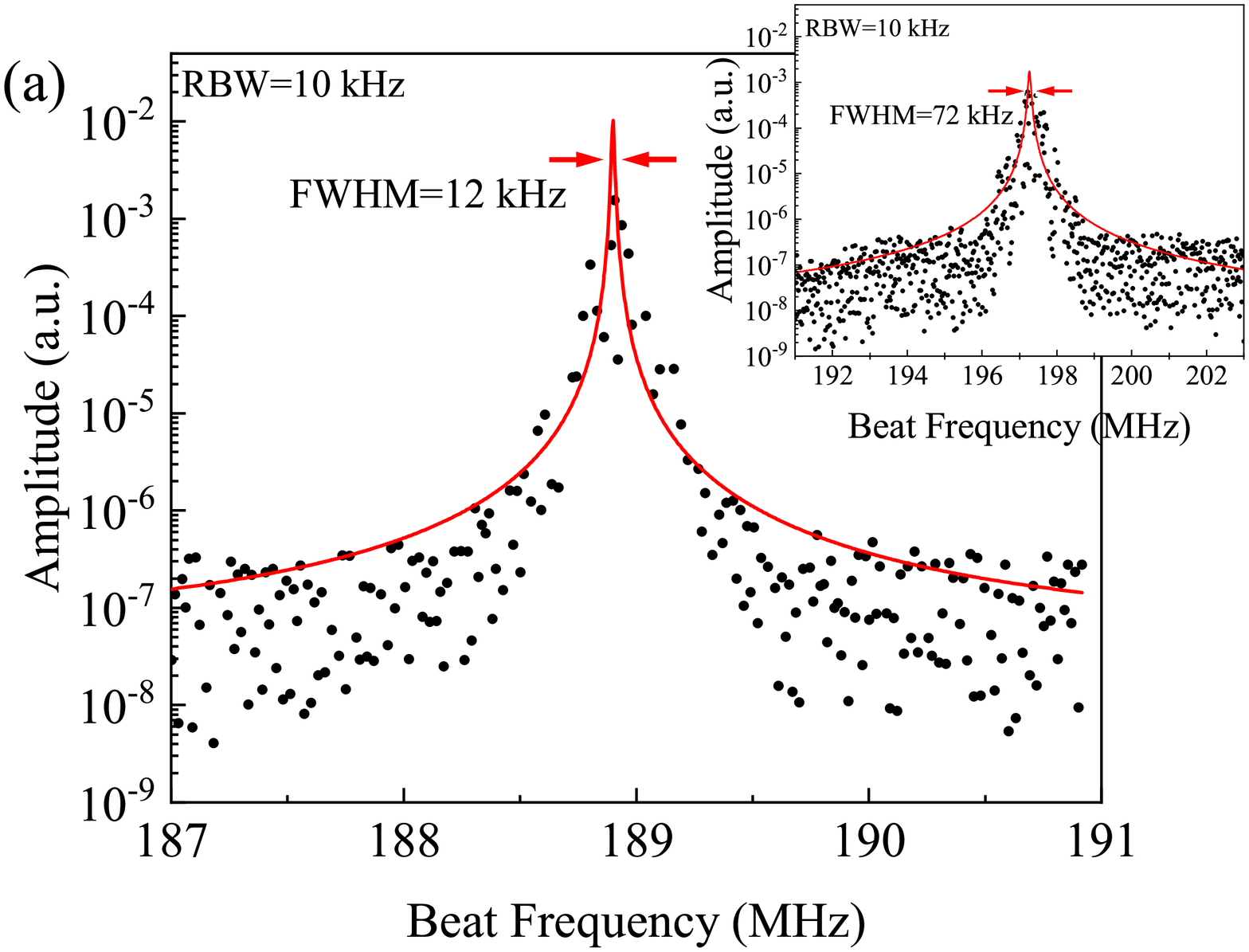}}
	\linebreak
	\subfigure{
		\label{histogram}
		\includegraphics[width=\linewidth]{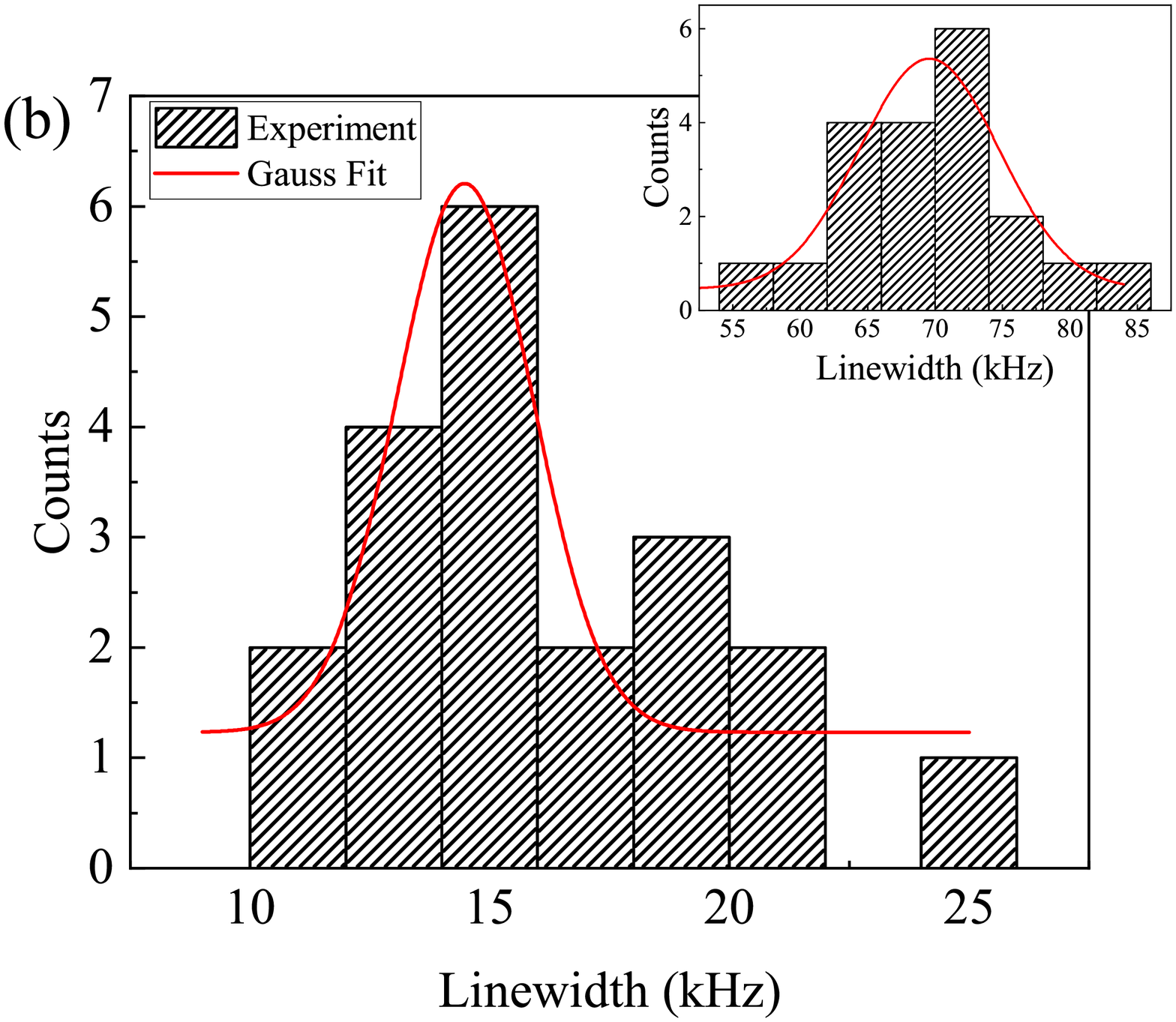}}
	\caption{(a) Typical beating data of two identical MTS-stabilized 459~nm lasers. The Lorentz-fitted FWHM of the beat linewidth is 12~kHz; Inset: typical beating data with one laser free-running, the fitted FWHM is 72~kHz.
	(b) Histogram of repeated measurements of the fitted beating linewidths with both lasers locked, the mean linewidth after locking is 14.5~kHz, indicating a linewidth of 10.3~kHz for each laser; Inset: histogram of repeated measurements of the fitted linewidths with one laser free-running, mean linewidth is 69.6 kHz. All linewidth measurements are made with the RBW of the spectrum analyzer set at 10~kHz, in order to have a fair comparison between the locked and free-running linewidths.}
\end{figure}

In addition to self-estimation evaluation, we measure the beat signal between two MTS systems with identical setups. One system is locked to the 6S$_{1/2}$ ({F}=4) - 7P$_{1/2}$ ({F'}=3) transition, with the other system locked to the 6S$_{1/2}$ ({F}=4) - 7P$_{1/2}$ ({F'}=3\&4) crossover transition. After combining the laser output of the two systems via a fiber-optical coupler, we use a high bandwidth photodetector (Hamamatsu C5658, bandwidth 1 GHz) to measure the beat signal with frequency equal to half of the frequency difference between the 7P$_{1/2}$ ({F'}=3) and 7P$_{1/2}$ ({F'}=4) hyperfine levels. 
The beat signal is sent into a spectrum analyzer (Keysight N9002B) for measuring the laser's linewidth. Results are shown in Figs.~\ref{linewidth} and~\ref{histogram}. After repeated measurements, the mean beating linewidth between two MTS-stabilized lasers is calculated to be 14.5~kHz, and the mean beating linewidth between a MTS-stabilized laser and a free-running laser is 69.6 kHz. Assuming equal contribution from each laser in the beating linewidth between two MTS-stabilized lasers, the linewidth of each laser is 10.3~kHz, whereas the beat linewidth between free-running and locked lasers mainly come from the free-running one. Therefore, the laser linewidth is reduced by a factor of 6.75 after locking. This reduction is due to suppression of the laser frequency noise by the MTS feedback~\cite{Ito2000}. 

The beat signal between two locked identical MTS systems is then sent into a frequency counter (Keysight 53230A) for measurement of frequency stability.
The Allan deviation can be calculated from the measured beat frequency sequence. The Allan deviation of the beatnote is $3\times{10}^{-13}/\sqrt{{\tau}}$, and assuming the two identical MTS systems contribute equally, the Allan deviation of each stabilized laser is $1/\sqrt{2}$ of the beat sequence, i.e. $2.1\times{10}^{-13}/\sqrt{{\tau}}$, as depicted by Fig.~\ref{Allan}.
The Allan deviation of the beatnote between a free-running laser and a locked laser is $8.2\times{10}^{-10}$ for averaging time below 10~s, and becomes $2\times{10}^{-10}\tau^{1/2}$ at higher averaging times.
The Allan deviation measured via heterodyne measurement is larger than that of the self-estimation (in-loop measurement), mostly due to out-of-loop elements such as temperature fluctuation of the cesium cells~\cite{Shang2020}.

\begin{figure}[htbp]
	\includegraphics[width=\linewidth]{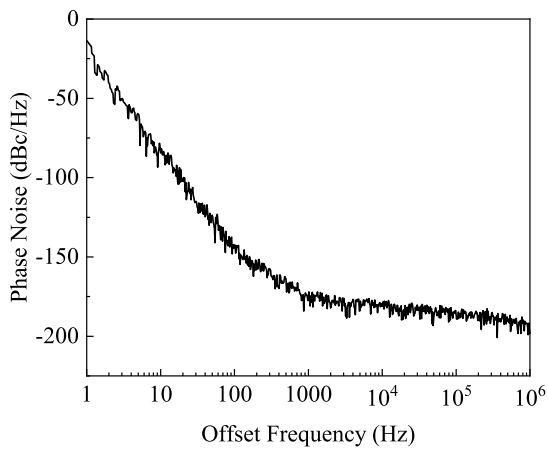}
	\caption{{\label{phase-noise}}Measured phase noise spectrum of the beat signal between two lasers locked to the $^{133}$Cs 6S$_{1/2}$ (F=4) -7P$_{1/2}$ (F'=3) transition and the F=4 - F'=4\&3 crossover transition, respectively.}
\end{figure}

We also measure the phase noise spectrum of the beat signal between two MTS-stabilized lasers with a phase noise analyzer (Keysight E5052B), and the result is shown in Fig.~\ref{phase-noise}.
However, since we only measured the phase noise spectrum in the frequency range of 1~Hz to 1~MHz, it corresponds to timescales less than 1~s, whereas our frequency stability measurement of the beatnote only starts at 0.1~s, so we can't make a direct comparison between the phase noise spectrum and frequency stability measurements. Instead, the phase noise spectrum supplements the frequency stability measurement and provides information on the beat signal's behavior at short timescales.

\subsection{Hyperfine level measurements with MTS}

\begin{figure}[htbp]
	\includegraphics[width=\linewidth]{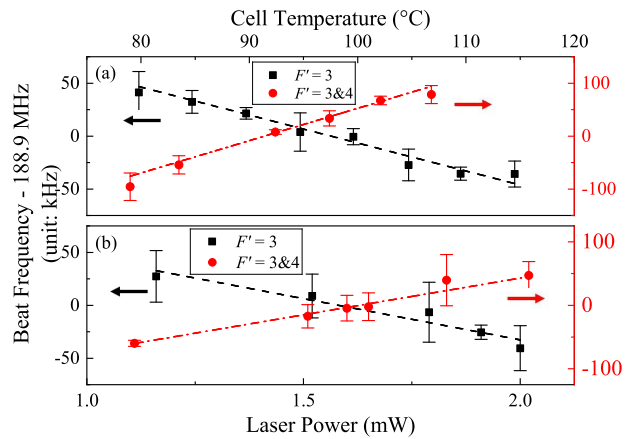}
	\caption{\label{shift}
		Experimental measurements of the beat frequencies between the Cs 6S$_{1/2}$~(F=4) - 7P$_{1/2}$~(F=3) transition (black squares) and the 6S$_{1/2}$~(F=4) - 7P$_{1/2}$~(F=3\&4) crossover transition (red circles): (a) under different cell temperatures; (b) under different total laser power. 
		Each data point in this figure is the averaged result from beat frequency data measured with a frequency counter with more than 300~s measurement time, and the error bars are the standard deviation calculated from each measurement result. The frequency shift coefficients are obtained with least-square fitting, and the linear fits are shown as dashed lines in the figure.}
\end{figure}

Since the beat frequency between the {F}=4~-~{F'}=3 transition and the {F}=4~-~{F'}=3\&4 crossover transition is equal to half of the hyperfine splitting of the 7P$_{1/2}$ energy level, the heterodyne measurement can also be used to measure the hyperfine structure of the 7P$_{1/2}$ energy level. By varying the cell temperature and the pump and probe laser power of one individual MTS system while keeping the other constant, the effects of collision shift and power shift can be accounted for and corrected. Figs.~\ref{shift}(a) and~\ref{shift}(b) show the relative frequency shifts of the {F}=4~-~{F'}=3 and crossover transitions under different cell temperatures and total laser power, and the shift coefficients are obtained via least-square fitting. The collision shift coefficients of the F=4 - F'=3 and crossover transitions are 2.64(17)~kHz/K and 6.11(43)~kHz/K, respectively. The ac Stark shift coefficients of the F=4 - F'=3 and crossover transitions are 77.89$\pm$9.81~kHz/mW and 115.65$\pm$4.54~kHz/mW, respectively.
Note that the frequency shift coefficients of the F=4 - F'=3 transition are of the opposite sign of the fitted slopes in Figs.~\ref{shift}(a) and~\ref{shift}(b), since increasing the F=4 - F'=3 transition frequency (therefore increasing the frequency of the laser locked to the transition) will decrease the beat frequency, and vice versa.
The hyperfine splitting can be calculated by extrapolating to null laser power and a cell temperature comparable with other experimental measurements, such as Ref.~\cite{RevModPhys.49.31} which uses optical double resonance (ODR) to measure the hyperfine splitting at room temperature. The calculated hyperfine splitting of the Cs 7P$_{1/2}$ energy level is 377.52$\pm$0.24~MHz. The hyperfine splitting from the transition's center of gravity is given by
\begin{equation}
	W=\frac{1}{2}AK+B\frac{(3/2)K(K+1)-2I(I+1)J(J+1)}{2I(2I-1)2J(2J-1)},
	\label{splitting}
\end{equation}
\noindent where $K=F(F+1)-I(I+1)-J(J+1)$; \textit{A} and \textit{B} are the magnetic dipole constant and the electric quadrupole constant, respectively; \textit{I}, \textit{J} and \textit{F} are the nuclear spin, the total electronic angular momentum, and the total atomic angular momentum, respectively. The hyperfine splitting between 7P$_{1/2}$~{F'}=3 and 7P$_{1/2}$~{F'}=4 equals to 4\textit{A}, and the measurement result corresponds to a magnetic dipole constant \textit{A} of 94.38(6)~MHz. As shown in Table~\ref{hyperfine}, our measurement result agree well with previous works, with a comparable level of measurement uncertainty~\cite{RevModPhys.49.31,Gerheardt1972,Williams_2018,PhysRevA.100.042506}. By using a different method to measure the hyperfine $A$ constant, our results provide an independent verification for previous works.

\begin{table}[bthp]
	\caption{\label{hyperfine}%
		Measured magnetic dipole constant $A$ of the Cs 7P$_{1/2}$ energy level, compared with earlier works.
	}
	\begin{ruledtabular}
		\begin{tabular}{llc}
			A(MHz)&Method&Reference\\
			\colrule\rule{0pt}{1.2em}%
			94.35(4)&ODR&\cite{RevModPhys.49.31}\\
			94.35(5)&SAS&\cite{Gerheardt1972}\\
			94.40(5)&SAS&\cite{Williams_2018}\\
			95.015&Theory&\cite{PhysRevA.100.042506}\\
			94.38(6)&MTS&This work\\
		\end{tabular}
	\end{ruledtabular}
\end{table}

\subsection{Noise sources in the MTS system}

Based on our measurements, we evaluate the major noise sources in our system and give a brief analysis on their impact on the performance of our compact optical clock.
Since the total measured frequency instability of $2.1\times{10}^{-13}/\sqrt{\tau}$, obtained by beating measurement, is an order of magnitude larger than the in-loop self-estimated frequency instability of $1.4\times{10}^{-14}/\sqrt{\tau}$, we mainly focus on the effects of out-of-loop elements that limit the frequency stability of the system. 

In atomic frequency standards based on vapor cells, collision between the thermal atoms introduce frequency shifts to the atomic energy levels. Since the collision shift depends on the atomic density, fluctuations of the cell temperature translates to fluctuations of the transition frequency. From the linear fitting in Fig.~\ref{shift}(a), the collision shift coefficient is measured to be 2.64(17)~kHz/K, or $\Delta f/f=(4.04\pm0.26)\times{10}^{-12}~\mathrm{K}^{-1}$ in terms of relative frequency shift. The cesium cell's temperature instability is estimated to be 0.05~K at 1~s, corresponding to a frequency instability of $2.02\times{10}^{-13}$.

External electromagnetic fields cause frequency shifts to the reference transition, and their fluctuations also introduce instabilities. We measure the B-field inside the two layers of mu-metal magnetic shielding to be approximately 1~mG. In a static magnetic field, the energy shift of the ground level can be calculated with the Breit-Rabi formula~\cite{PhysRev.38.2082.2}, which yields a quadratic shift of 213.7 Hz/G$^2$ for the 6S$_{1/2}$ F=4 hyperfine level. As fort the 7P$_{1/2}$ F=3 excited state, we make an estimate based on the fact that, the Zeeman shift of the $m_F=0$ sublevel must be smaller than the splitting between adjacent sublevels in the 7P$_{1/2}$ F=3 level~\cite{Shang2020}. We therefore estimate the Zeeman shift of the excited level to be below 466~kHz/G. With the B-field below 1~mG, the total Zeeman shift of the reference transition is below 46~Hz. Assuming the variation of the magnetic field to be less than 10\%, the frequency instability induced by the Zeeman effect is estimated to be below $7\times{10}^{-15}$.

Meanwhile, due to ac Stark effect, any fluctuation in the laser power is translated to fluctuation of the reference transition's frequency shift. Similar to the above, we calculate the relative frequency shift to be $\Delta f/f=(1.19\pm0.15)\times{10}^{-10}~\mathrm{mW}^{-1}$ from the measured ac Stark shift under different laser power, as shown in Fig.~\ref{shift}(b). The measured laser power instability is $1.2\times{10}^{-4}$ at 1~s, and the total laser power is about 1.83~mW (pump laser power 1.5~mW and probe laser power 0.33~mW), so the laser power instability corresponds to $2.6\times{10}^{-14}$ frequency instability at 1~s.

The combined effect of cell temperature fluctuation, external magnetic field, and laser power fluctuation produce a frequency instability of $2.04\times{10}^{-13}$ at 1~s, which is in good agreement with the measured short-term frequency of $2.1\times{10}^{-13}/\sqrt{\tau}$. Therefore, we conclude that the main limitation to the short-term frequency stability is the temperature fluctuation of the cesium cell.
In future works, we plan to utilize a double-layered atomic cell with vacuum heat insulation in conjunction with a better temperature-control system, so as to reduce the temperature fluctuation of the atomic ensemble and thereby improve the frequency stability of the MTS-stabilized laser.

\section{Conclusion}
In this study, we investigate a compact optical frequency standard based on modulation transfer spectroscopy of thermal cesium 133 atoms. By optimizing the operating parameters of the MTS stabilization, we realize a self-estimated short-term stability of $1.4\times{10}^{-14}/\sqrt{\tau}$, reaching $4\times{10}^{-15}$ at 30~s. We measure the linewidth and Allan deviation via heterodyne, verifying the linewidth-narrowing effect of MTS locking, with the mean beating linewidth reduced from 69.6~kHz to 10.3~kHz, by a factor of 6.75.
The Allan deviation of the MTS-stabilized laser is calculated from the beat frequency sequence to be $2.1\times{10}^{-13}/\sqrt{\tau}$.
By measuring the beat frequency between two MTS systems locked to the 6S$_{1/2}$ ({F}=4) - 7P$_{1/2}$ ({F'}=3) transition and the 6S$_{1/2}$ ({F}=4) - 7P$_{1/2}$ ({F'}=3\&4) crossover transitions respectively, we also measure the hyperfine structure of the 7P$_{1/2}$ energy level,  and calculate the magnetic dipole constant \textit{A} of the Cs 7P$_{1/2}$ energy level to be 94.38(6)~MHz, which agrees with previous works~\cite{RevModPhys.49.31,Gerheardt1972,Williams_2018,PhysRevA.100.042506}.
In future works, we expect to improve the long-term frequency stability of this optical frequency standard by utilizing a double-layered atomic cell with vacuum heat insulation in conjunction with a better temperature-control system to reduce the temperature fluctuation of the atomic ensemble, as well as actively controlling the temperature of the EOM to reduce the residual amplitude modulation, both of which are limiting factors on the optical frequency standard's long-term stability.
In addition to being used as an optical frequency standard, this compact, narrow-linewidth, high-stability laser may find various other applications, such as laser interferometry, laser cooling, geodesy, and so on.


\bibliography{459MTS-PRApplied}

\end{document}